\def\fracm#1#2{\hbox{\large{${\frac{{#1}}{{#2}}}$}}}

\documentstyle[12pt]{article}

\def\@magscale#1{ scaled \magstep #1}

\catcode`@=11  
\def\un#1{\relax\ifmmode\@@underline#1\else
        $\@@underline{\hbox{#1}}$\relax\fi}
\catcode`@=12

\def\a{\alpha}
\def\b{\beta}

\def\d{\delta}

\def\p{\pi}
\def\q{\theta}

\def\s{\sigma}

\def\G{\Gamma}

\def\L{\Lambda}

\def\P{\Pi}

\def\dslash{\not{\hbox{\kern-2pt $\partial$}}}
\def\Dslash{\not{\hbox{\kern-4pt $D$}}}
\def\pslash{\not{\hbox{\kern-2.3pt $p$}}}
 \newtoks\slashfraction
 \slashfraction={.13}
 \def\slash#1{\setbox0\hbox{$ #1 $}
 \setbox0\hbox to \the\slashfraction\wd0{\hss \box0}/\box0 }
 
\font\ro=cmsy10                          
\def\kcr{{\hbox{\ro \char'170}}}                
\def\ktl{{\hbox{\ro \char'170}}}        
\def\ktr{{\hbox{\ro \char'170}}}        
\def\kbl{{\hbox{\ro \char'170}}}        
\def\kbr{{\hbox{\ro \char'170}}}        

\def\plpl{\raise-2pt\hbox{$\raise3pt\hbox{$_+$}\hskip-6.67pt\raise0.0pt
\hbox{$^+$}\hskip 0.01pt$}}
\def\mimi{\raise-2pt\hbox{$\raise3pt\hbox{$_-$}\hskip-6.67pt\raise0.0pt
\hbox{$^-$}\hskip 0.01pt$}} 

\def\bo{{\raise.15ex\hbox{\large$\Box$}}}               
\def\pa{\partial}                                       
\def\TH{{\raise.2ex\hbox{$\displaystyle \bigodot$}\mskip-4.7mu \llap H \;}}
\def\face{{\raise.2ex\hbox{$\displaystyle \bigodot$}\mskip-2.2mu \llap {$\ddot
        \smile$}}}                                      

\def\sp#1{{}^{#1}}                              
\def\Tilde#1{\widetilde{#1}}                    
\def\Hat#1{\widehat{#1}}                        
\def\Bar#1{\overline{#1}}                       
\def\leftrightarrowfill{$\mathsurround=0pt \mathord\leftarrow \mkern-6mu
        \cleaders\hbox{$\mkern-2mu \mathord- \mkern-2mu$}\hfill
        \mkern-6mu \mathord\rightarrow$}
\def\dvec#1{\vbox{\ialign{##\crcr
        \leftrightarrowfill\crcr\noalign{\kern-1pt\nointerlineskip}
        $\hfil\displaystyle{#1}\hfil$\crcr}}}           

\def\fracm#1#2{\hbox{\large{${\frac{{#1}}{{#2}}}$}}}
\def\frac#1#2{{\textstyle{#1\over\vphantom2\smash{\raise.20ex
        \hbox{$\scriptstyle{#2}$}}}}}                   
\def\sfrac#1#2{{\vphantom1\smash{\lower.5ex\hbox{\small$#1$}}\over
        \vphantom1\smash{\raise.4ex\hbox{\small$#2$}}}} 
\def\bfrac#1#2{{\vphantom1\smash{\lower.5ex\hbox{$#1$}}\over
        \vphantom1\smash{\raise.3ex\hbox{$#2$}}}}       
\def\afrac#1#2{{\vphantom1\smash{\lower.5ex\hbox{$#1$}}\over#2}}    

\newskip\humongous \humongous=0pt plus 1000pt minus 1000pt
\def\caja{\mathsurround=0pt}
\def\eqalign#1{\,\vcenter{\openup2\jot \caja
        \ialign{\strut \hfil$\displaystyle{##}$&$
        \displaystyle{{}##}$\hfil\crcr#1\crcr}}\,}
\newif\ifdtup

\def\ref#1{$\sp{#1)}$}

\parskip=\medskipamount                 
\thicklines                         

\def\oldheadpic{                                
        \setlength{\unitlength}{.4mm}
        \thinlines
        \par
        \begin{picture}(349,16)
        \put(325,16){\line(1,0){4}}
        \put(330,16){\line(1,0){4}}
        \put(340,16){\line(1,0){4}}
        \put(335,0){\line(1,0){4}}
        \put(340,0){\line(1,0){4}}
        \put(345,0){\line(1,0){4}}
        \put(329,0){\line(0,1){16}}
        \put(330,0){\line(0,1){16}}
        \put(339,0){\line(0,1){16}}
        \put(340,0){\line(0,1){16}}
        \put(344,0){\line(0,1){16}}
        \put(345,0){\line(0,1){16}}
        \put(329,16){\oval(8,32)[bl]}
        \put(330,16){\oval(8,32)[br]}
        \put(339,0){\oval(8,32)[tl]}
        \put(345,0){\oval(8,32)[tr]}
        \end{picture}
        \par
        \thicklines
        \vskip.2in}
\def\oldtitle#1#2#3#4{\oldheadpic\begin{center}\vglue.5in{\large\bf #1}\\[.6in]
        {#2}\\[.1in] {\it Department of Physics and Astronomy}\\
        {\it University of Maryland, College Park, MD 20742}\\[.6in]
        Physics Publication \#{#3}\\ {#4}\\[1.5in] {\bf ABSTRACT}\\[.1in]
        \end{center} \begin{quotation}}                 
\def\oldTitle#1#2#3#4#5#6#7{\oldheadpic\begin{center} \vglue .4in
        {\large\bf #1}\\[.4in]
        {#2}\\[.1in] {\it Department of Physics and Astronomy}\\
        {\it University of Maryland, College Park, MD 20742}\\[.1in]
        {#3}\\[.1in] {\it {#4}}\\ {\it {#5}}\\[.4in]
        Physics Publication \#{#6}\\ {#7}\\[.5in] {\bf ABSTRACT}\\[.1in]
        \end{center} \begin{quotation}}                 
\def\border{                                            
        \setlength{\unitlength}{1mm}
        \newcount\xco
        \newcount\yco
        \xco=-21
        \yco=12
        \begin{picture}(140,0)
        \put(\xco,\yco){$\ktl$}
        \advance\yco by-1
        {\loop
        \put(\xco,\yco){$\kcr$}
        \advance\yco by-2
        \ifnum\yco>-240
        \repeat
        \put(\xco,\yco){$\kbl$}}
        \xco=158
        \yco=12
        \put(\xco,\yco){$\ktr$}
        \advance\yco by-1
        {\loop
        \put(\xco,\yco){$\kcr$}
        \advance\yco by-2
        \ifnum\yco>-240
        \repeat
        \put(\xco,\yco){$\kbr$}}
        \put(-20,13){\tiny University of Maryland Elementary Particle
Physics University of Maryland Elementary Particle Physics University of
Maryland Elementary Particle Physics}
        \put(-20,-241.5){\tiny University of Maryland Elementary
Particle Physics University of Maryland Elementary Particle Physics
University of Maryland Elementary Particle Physics}
        \end{picture}
        \par\vskip-8mm}
\def\bordero{                                           
        \setlength{\unitlength}{1mm}
        \newcount\xco
        \newcount\yco
        \xco=-31
        \yco=12
        \begin{picture}(140,0)
        \put(\xco,\yco){$\ktl$}
        \advance\yco by-1
        {\loop
        \put(\xco,\yco){$\kclr}
        \advance\yco by-2
        \ifnum\yco>-240
        \repeat
        \put(\xco,\yco){$\kbl$}}
        \xco=151
        \yco=12
        \put(\xco,\yco){$\ktr$}
        \advance\yco by-1
        {\loop
        \put(\xco,\yco){$\kcr$}
        \advance\yco by-2
        \ifnum\yco>-240
        \repeat
        \put(\xco,\yco){$\kbr$}}
        \put(-20,12){\ooo bacdefghidfghghdhededbihdgdfdfhhdheidhdhebaaahjhhdahba

hgdedge
   hgfdiehhgdigicba}
        \put(-20,-241.5){\ooo ababaighefdbfghgeahgdfgafagihdidihiidhiagfedhadbfd

ecdcdfa
   gdcbhaddhbgfchbgfdacfediacbabab}
        \end{picture}
        \par\vskip-8mm}
\def\headpic{                                           
        \indent
        \setlength{\unitlength}{.4mm}
        \thinlines
        \par
        \begin{picture}(29,16)
        \put(165,16){\line(1,0){4}}
        \put(170,16){\line(1,0){4}}
        \put(180,16){\line(1,0){4}}
        \put(175,0){\line(1,0){4}}
        \put(180,0){\line(1,0){4}}
        \put(185,0){\line(1,0){4}}
        \put(169,0){\line(0,1){16}}
        \put(170,0){\line(0,1){16}}
        \put(179,0){\line(0,1){16}}
        \put(180,0){\line(0,1){16}}
        \put(184,0){\line(0,1){16}}
        \put(185,0){\line(0,1){16}}
        \put(169,16){\oval(8,32)[bl]}
        \put(170,16){\oval(8,32)[br]}
        \put(179,0){\oval(8,32)[tl]}
        \put(185,0){\oval(8,32)[tr]}
        \end{picture}
        \par\vskip-6.5mm
        \thicklines}
\def\title#1#2#3#4{\border\headpic {\hbox to\hsize{#4 \hfill UMDEPP #3}}\par
        \begin{center} \vglue .5in {\large\bf #1}\\[.6in]
        {#2}\\[.1in] {\it Department of Physics and Astronomy}\\
        {\it University of Maryland, College Park, MD 20742}\\[1.5in]
        {\bf ABSTRACT}\\[.1in] \end{center} \begin{quotation}}  
\def\Title#1#2#3#4#5#6#7{\border\headpic
        {\hbox to\hsize{#7 \hfill UMDEPP #6}}\par
        \begin{center} \vglue .4in {\large\bf #1}\\[.4in]
        {#2}\\[.1in] {\it Department of Physics and Astronomy}\\
        {\it University of Maryland, College Park, MD 20742}\\[.1in]
        {#3}\\[.1in] {\it {#4}}\\ {\it {#5}}\\[.5in] {\bf ABSTRACT}\\[.1in]
        \end{center} \begin{quotation}}                 
\def\endtitle{\end{quotation}\newpage}                  

\def\ad{{\kern0.5pt
                   \alpha \kern-5.05pt \raise5.8pt\hbox{$\textstyle.$}\kern
0.5pt}}

\def\bd{{\kern0.5pt
                   \beta \kern-5.05pt \raise5.8pt\hbox{$\textstyle.$}\kern
0.5pt}}

\def\qd{{\kern0.5pt
                   q \kern-5.05pt \raise5.8pt\hbox{$\textstyle.$}\kern
0.5pt}}
\def\Dot#1{{\kern0.5pt
                   {#1} \kern-5.05pt \raise5.8pt\hbox{$\textstyle.$}\kern
0.5pt}}

\begin{document}

\def\gfrac#1#2{\frac {\scriptstyle{#1}}
        {\mbox{\raisebox{-.6ex}{$\scriptstyle{#2}$}}}}
\def\gg{{\hbox{\sc g}}}
\border\headpic {\hbox to\hsize{December 2000 \hfill {Bicocca--FT--00--26}}}
\par \hfill {BRX TH--485}
\par \hfill {McGill 00--34}
\par \hfill {UMDEPP 00--144}
\par
\setlength{\oddsidemargin}{0.3in}
\setlength{\evensidemargin}{-0.3in}
\begin{center}
\vglue .05in
{\Large\bf The Superspace WZNW Action \\[.09in]
for 4D, N=1 Supersymmetric QCD\footnote {
Supported in part by National 
Science Foundation Grants PHY-98-02551, PHY-00-70475, by MURST
and European Commission TMR program ERBFMRX-CT96-0045 in which S.P. is 
associated to the University of Padova.}  }
\\[.05in]
S. James Gates, Jr.\footnote{gatess@wam.umd.edu}
\\[0.02in]
{\it Department of Physics, University of Maryland\\ 
College Park, MD 20742-4111 USA}
\\[.05in] 
Marcus T. Grisaru, \footnote{grisaru@brandeis.edu}
\\[0.02in]
{\it Physics Department, Brandeis University \\
Waltham, MA 02454, USA}
\\[.05in]
Marcia E. Knutt, \footnote{knutt@physics.mcgill.ca}
\\[0.02in]
{\it Physics Department, McGill University \\
Montreal, QC Canada H3A 2T8}\\[0.02in]
and \\ [.02in] Silvia Penati\footnote{Silvia.Penati@mi.infn.it}
\\[0.02in]
{\it Dipartmento di Fisica dell'Universit\' a di 
Milano--Bicocca \\
and INFN, Sezione di Milano, p.za della Scienza 3, 
I-20126 Milano, Italy}
\\[.06in]
{\bf ABSTRACT}\\[.002in]
\end{center}
\begin{quotation}
{We discuss features of the 4D, $N$ = 1 WZNW term expressed
as a function of chiral superfields and defined in a manner 
appropriate to calculate phenomenological matrix elements.}

${~~~}$ \newline


\endtitle

\noindent
{\Large {\bf {(I.) Introduction}}  }

A possible interpretation of the recent experimental observation 
of a ``light'' Higgs particle\cite{HGSX}  at 114.5 GeV is that 
the probability of finding supersymmetry in Nature   grows ever 
larger.  It is, therefore, appropriate  that techniques be 
developed and results worked out with an eye toward developing
(in principle at least) experimentally verifiable tests.  One 
place in which this  {\it {might}} be done is within the context  
of effective actions for 4D, $N$ = 1 SUSY QCD.  This is an area 
that has attracted our attention during the course of the past 
few years \cite{Nonmwznw}. We have explored the possibility 
of a non-conventional description of the 4D, $N$ = 1 supersymmetric 
WZNW action and its possible role as an effective action for 
low-energy 4D, $N$ = 1 SUSY QCD; we have also investigated 
the more conventional description and within this approach 
written the first practical description \cite{GGP,GGKPS} of 
the BGJ anomaly action and the corresponding expression for 
the gauged 4D, $N$ = 1 supersymmetric WZNW action \cite{GGP}.   
In this paper we discuss some features of this latter result.

For many years a treatment of the 4D, $N$ = 1 supersymmetric WZNW 
action given in the literature by Nemeschansky and Rohm (N-R) 
\cite{NR} has been the standard basis for considering properties
of the low-energy 4D, $N$ = 1 SUSY QCD effective action related
to the anomaly (see for example \cite{Man}).  This  description 
possesses an infinite number of unspecified constants that appear 
in an undetermined function (denoted by $\b_{{i} {j} {\Bar {k}}}
$) in the N-R work\footnote{Even prior to its publication, the 
N-R action was in the unpublished notes of one of the \newline 
${~~~~\,}$ present authors (SJG) where it remained  due to the 
lack of an explicit expression for\newline ${~~~~\,}$ the
$\b$-coefficient.}.   These  constants enter in the calculations 
of matrix elements based upon the N-R WZNW action.  Any physically 
testable process described  by the N-R WZNW model is largely 
unspecified due to this.  In the N-R work \cite{NR} a discussion 
was given of complex geometry, Stein manifolds and the relevant 
Dolbeault cohomology groups related to  a tensor such as $\b_{
{i} {j} {\Bar {k}}}$.  Unfortunately, since this tensor was not 
determined, one could not derive specific physical consequences or 
predictions from the N-R action.

On the other hand, our recent work \cite{GGP} based on the minimal 
homotopy of 4D, $N$ = 1 supersymmetric Yang-Mills theory \cite{GGP,
GGKPS} is quite predictive.   We have presented  in these references 
the entire non-linear and therefore non-perturbative structure of 
the 4D, $N$ = 1 anomaly and  WZNW term.  We will use these new results 
to explore the formulation of the WZNW terms in the low-energy 
effective 4D, $N$ = 1 supersymmetric QCD action.

\newpage

\noindent
{\Large {\bf {(II.) Explicit 4D, $N$ = 1 Superspace WZNW Action
}}}

${~~}$ \newline \indent
In previous work \cite{GGP, GGKPS} it was shown that the consistent 
anomaly in 4D, $N$ = 1 supersymmetric Yang-Mills theories can be cast 
in the form
$$
{\cal A}_{BGJ}( \L\,; e^V) ~=~ \fracm 1{ \, 4 \pi^2 \,  } 
\,{\cal I
}{\rm {m}} \Big[ ~ \int d^4 x ~ d^2 \q ~ d^2 {\bar \q} ~ {\cal P}( 
\L\,; e^V) \, \Big] ~~~, {~~~~~~~~~~~~~~~~~~~~~~~~} \eqno(2.1) 
$$
where
$$
\eqalign{  {~~~~}
{\cal P}( \L\,; e^V) ~=~  {\rm Tr} \Big[ \, \L \, \Big( 
\, \G^{\a} W_{\a} \,- \, \int_0^1 \, dy \, y ~ ( ~ &[ \, {\cal 
W}^{\a} \, , \, \p_{\a} \, ] \, e^V  \, {\cal G}{~~~~~~~~~}  \cr
&+~ \{ \, \Tilde{\cal W}^{\dot \a} \, ,~ {\bf I} \,- \, e^V \, 
{\cal G} \,\} \, \Tilde{\p}_{\dot \a}  ~)~ \Big)~ \Big]   
~~~ {~~}} 
\eqno(2.2) $$
is a function of the chiral gauge parameter $\L$ and of the following
geometric objects:
$$ \eqalign{ {~~~~~~~}
{\cal W}_{\a} &\equiv~ i \,  {\Bar D}{}^2 ( {\cal G} D_{\a} {\cal 
G}^{-1})  ~~~~~~,~~~ {\Tilde {\cal W}}{}_{\Dot \a} ~ \equiv~ i
\,{\cal G} \{ D{}^2 ({\cal G}^{-1} \, 
{\Bar D}{}_{\Dot \a}{\cal G} ) \, \}  {\cal G}^{-1} \,~~~, \cr 
\p_{\a} &\equiv~ {\cal G}^2 \, e^V \, \G_{\a}  {~~~~~~~~~~~~~\,} ,  
{~\,~~~} \Tilde{\p}_{\dot \a} ~\equiv~ e^V \, {\cal G} \, \Tilde{\G
}_{\dot \a} \, {\cal G} ~~~, {~~~~~~} \cr 
\G_{\a} &\equiv~ i \, e^{-V} \, D_{\a} e^V {~~~~~~~~~~\,} ,  
{~\,~~~} \Tilde{\G}_{\dot \a} ~\equiv~ e^{-V} \, \bar{\G}_{
\dot \a} \, e^V ~~~, {~~~~~~} \cr 
{\cal G} &\equiv~ \Big[ ~ {\bf I} ~+~ y \, (\, e^V \, - \, {\bf I} 
\,) ~ \Big]^{-1}
~~~.}
\eqno(2.3) $$

A set of parametric gauge orbits of the Yang-Mills superfields which
appear in this action are defined by the equations
$$
\eqalign{ {~}
\G^{\prime}{}_{\a} &\equiv~ {{} U}{}^{-1} \, \G^{}{}_{\a} \, {{} 
U} ~+~ i \, {{} U}{}^{-1} \, D_{\a} \, {{} U} ~~~~~\,, ~~ W^{\prime
}{}_{\a} ~\equiv~ {{} U}{}^{-1} \, W_{\a} \, {{} U} ~~,~~ \cr 
e^{\, V^{\prime} \,} &\equiv~ {{} U}{}^{\dagger} \, e^{\, V \,} ~ 
{{} U} ~~~, ~~~ {{} U} ~\equiv~ e^{ - i w {\L}/ f } ~~~,~~~ {
\Bar D}{}_{\ad} \, \L ~=~ 0  ~~~.
} 
\eqno(2.4) $$
We have introduced a constant $f$ with the dimensions of mass so as 
to allow $\L$ to have canonical engineering dimensions.  It thus 
follows that the 4D, $N$ = 1 supersymmetric gauged WZNW term takes 
the form\footnote{The second line of (2.5) corrects  a misprint in 
the corresponding  expression in \cite{GGP}.}${}^,$\footnote{Since ultimately it is our desire for this action
to be as close as possible to the form of a use- \newline ${~~~~\,}$ ful
phenomenological action, we have introduced a normalization constant
$C_0/f$ to multiply \newline ${~~~~\,}$ the result of our previous work.
The actual value of the dimensionless constant $C_0$ is determin-
\newline ${~~~~\,}$ ed by looking at the purely bosonic limit of this
action.}
$$ \eqalign{ {~}
{\cal S}^{gauged}_{WZNW}( \L\,; e^{V^{\prime}}) &=~
C_0 \, (\fracm 1{ \, 4 \pi^2 \, f \, } )
\,{\cal I}{\rm {m}} \Big[ ~ \int d^4 x \, d^2 \q \, d^2 
{\bar \q} \int_0^1 d w ~ {\cal P}( \L\,; {{} U}{}^{\dagger} \, 
e^{\, V \,} ~  {{} U}) \, \Big] \cr
&=~ C_0 \, (\fracm 1{ \, 4 \pi^2 \,  })
\, {\cal R}{\rm e} \Big[ ~ \int d^4 x \, 
d^2 \q \, d^2 {\bar \q} \int_0^1 d w ~ {\cal P}( {{} U}{}^{-1} \, 
\pa_w {{} U}{} \,; {{} U}{}^{\dagger} \, e^{\, V \,} ~ {{} U}) 
\, \Big]  ~~~.}
\eqno(2.5) $$

In the remainder of this work we only consider the ungauged version 
of this action obtained by setting the Yang-Mills superfield to zero
(i.e. evaluate the anomaly on a pure gauge orbit)
$$
{\cal S}{}_{WZNW}({{} U}) ~=~ C_0 \, (\fracm 1{\, 4 \pi^2 \,} )\, {\cal 
R}{\rm e} \Big[ ~ \int d^4 x \, d^2 \q \, d^2 {\bar \q} \int_0^1
d w \, \int_0^1 d y \, y ~ {\cal Q}( {{} U}) ~ \Big] ~~~,
\eqno(2.6) $$
where the integrand takes the form
$$ \eqalign{ {~~}
{\cal Q}({{} U}) ~=~ {\rm Tr} \Big[ ~ {{} U}{}^{-1}( \pa_w {{} 
U} ) \, \Big( \, &[ \, {\cal V}^{\a} \, , \, u_{\a} \, ] \, {{} 
U}{}^{\dagger} \, {{} U} \, {\Hat {\cal G}} ~+~ \{ \, \Tilde{{} 
\cal{V}}^{\dot \a} \, ,~ {\Hat {\cal H}} \,\} \, \Tilde{u}_{\dot \a}  
~ \Big)~ \Big]    ~~~, } \eqno(2.7) $$
$$ \eqalign{ {~}
{\cal V}_{\a} &\equiv~ i\, \,y \, {\Bar D}{}^{2} \Big( \, {\Hat 
{\cal G}} {{} U}^{\dagger} {{} U}\, {\mit\P}^L_{\a}  \, \Big)  
~~~~,~~ {\Tilde {\cal V}}{}_{\Dot \a} ~\equiv~  - \, i \,y \, 
{\Hat  {\cal G}}  \, D{}^{2} \Big( \,{\Bar {\mit\P}}^L_{\ad} \, 
{{} U}^{\dagger} {{} U} \, {\Hat  {\cal G}} \, \Big)  \, {\Hat 
{\cal G}}^{-1} ~~~, ~~~  \cr u_{\a} &\equiv~ i \, {\Hat {\cal 
G}}^2 \, {{} U}{}^{\dagger} {{}  U} \, {\mit\P}^L_{\a} 
{~~~~~\,~~~~~~~,~~\,} 
\Tilde{u}_{\dot \a} ~\equiv~ - i {\Hat {\cal G}}\, {\Bar
{\mit\P}}^L_{\ad}\, {\Hat {\cal G}} 
\, {{} U}{}^{\dagger} {{} U} ~~~, {~~~~~~} \cr  
{\Hat {\cal G}} &\equiv~ \Big[ ~ {\bf I} ~+~ y \, (\, {{} U}{}^{
\dagger} \, {{} U}  \, - \, {\bf I} \,) ~ \Big]^{-1} ~~~,~~~ {
\Hat {\cal H}} ~\equiv~ 1 \,- \, {{} U}{}^{\dagger} \, {{} 
U} \, {\Hat {\cal G}}  ~~~.}
\eqno(2.8)$$
We have defined
$$
{\Tilde {\cal V}}{}_{\ad} ~=~ {\Hat {\cal G}} \, (-{\cal V}{}_{\a}
)^{\dagger}\, {\Hat {\cal G}}{}^{-1} ~~~,~~~ {\Tilde u}{}_{\ad} 
~=~ {\Hat {\cal G}} \, (-u_{\a} )^{\dagger}\, {\Hat {\cal G}}{
}^{-1} ~~~, \eqno(2.9) $$
where the $\dagger$-operation involves superspace conjugation. 

The superfield $\Hat {\cal G}$ is the vanishing gauge superfield 
limit of the minimal homotopy $\cal G$.  In the above equations
we have also introduced the left-and right-invariant Maurer-Cartan 
forms as well as their hermitian conjugates via the definitions
$$ \eqalign{ {~~~}
{{} U}^{-1} d {{} U} ~\equiv~ {\mit\P}^L ~~~,~~~ [d {{} U}^{\dagger}
]\, [{{} U}^{\dagger}]^{-1} ~\equiv~ {\Bar {\mit\P}}{}^L ~~~, \cr
( d {{} U}) \,{{} U}^{-1} ~\equiv~ {\mit\P}^R ~~~,~~~ [{{} U}^{
\dagger}]^{-1} [d {{} U}^{\dagger}] ~\equiv~  {\Bar {\mit\P}}{}^R 
~~~,}
\eqno(2.10) $$
for any derivative operator $d$.  So for example, ${\mit\P}^L_{\a} = {{} 
U}^{-1} D_{\a} {{} U}$, etc.  The results in (2.6 - 2.8) define
an explicit 4D, $N$ = 1 supersymmetric WZNW action with two free 
parameters.  The quantities ${\cal V}_{\a}$ and ${\Tilde {\cal V
}}{}_{\Dot \a}$ can  be further expanded by using
$$ \eqalign{ {~~~~~~~~}
D_{\a} {\Hat {\cal G}} &=~ - y \, {\Hat {\cal G}} ~ [ ~ {{} U}
{}^{\dagger} {{} U} \, {\mit\P}^L_{\a} ~ ] \, {\Hat {\cal G}}  
~~~, ~~~ {\Bar D}{}_{\ad} {\Hat {\cal G}} ~=~ - y \, {\Hat {
\cal G}} ~ [ ~ {\Bar {\mit\P}}{}^L_{\ad} \, {{} U}{ }^{\dagger} 
{{} U}  ~ ] \, {\Hat {\cal G}}  ~~~, \cr
\pa_{\un a} {\Hat {\cal G}} &=~ - y \, {\Hat {\cal G}} ~ [ ~ 
{\Bar {\mit\P}}{}^L_{\un a} \, {{} U}{}^{\dagger} {{} U} ~+~ \, 
\, {{} U}{ }^{\dagger} {{} U} \, {\mit\P}^L_{\un a} ~ ] \, {\Hat 
{\cal G}} ~~~. }\eqno(2.11) $$
This leads to 
$$ \eqalign{ {~}
 {\cal V}{}_{\a} ~&=~  - \,y (\, 1 - y \,) \, \Big[ ~ {\Hat {\cal 
G}} \, {\Bar {\mit\P}}{}^{\,\ad} \, {{} U}{}^{\dagger} {{} U} 
\Hat{\cal G}\, {\mit\P}_{\un a}  ~-~ i \,  \fracm 12 \,  {\Hat 
{\cal G}} \, \Big( {\Bar D}{}^{\ad}  ~ {\Bar {\mit\P}}{}_{\,\ad}
\, -\, {\Bar {\mit\P}}{}^{\ad} {\Bar {\mit\P}}{}_{\,\ad} ~ \Big) 
\, {{} U}{}^{\dagger} {{} U} \Hat{\cal G}\, {\mit\P}_{\a}  \cr  
&{~~~~~~~~~~~~~~~~~~~~~} -~i  \, (1~-~y) {\Hat {\cal G}} \, {\Bar 
{\mit\P} }{}^{\ad}  \, {\Hat {\cal G}} \, {\Bar {\mit\P}}{}_{\,\ad}  
\, {{} U}{}^{\dagger} {{} U} \Hat{\cal G}\, {\mit\P}_{\a}\, 
\Big] ~~~, }   \eqno(2.12) $$
\vspace{0.1in}
$$ \eqalign{ {~~\,\,}
{\Tilde {\cal V}}{}_{\ad} ~&= y \,(\, 1 - y \,) \, \Big[ ~ {\Hat 
{\cal G}} \, {\Bar {\mit\P}}{}_{\un a} \, {{} U}{}^{\dagger} {
{} U} \Hat{\cal G} \, {{\mit\P}}{}^{\,\a}  ~-~i \fracm 12 \, 
{\Hat {\cal G}} \, {\Bar {\mit\P}}{}_{\,\ad} \, {{} U}{}^{\dagger} 
{{} U} \Hat{\cal G}\, \Big( {D}{}^{\a} ~ {{\mit\P}}{ }_{\,\a}\,-\,
{{\mit\P}}{}^{\a} {{\mit\P}}{}_{\,\a} ~ \Big) \cr
&{~~~~~~~~~~~~~~~~~~~}-~i  \,(1~-~y) \, {\Hat {\cal G}} \, {\Bar 
{\mit\P}}{}_{\ad} \, {{} U}{}^{\dagger} {{} U} \Hat{\cal G}\, 
{{\mit\P}}{}^{\,\a} \, \Hat{\cal G}{\mit\P}_{\a}  \, \Big] 
~~~.}   \eqno(2.13) $$
All of the superfield Maurer-Cartan forms in (2.12, 2.13) are
left-invariant forms but we have dropped the superscripts $L$. 

Expressed in terms of Maurer-Cartan forms and their derivatives, 
the function ${\cal Q}({{} U})$ is given by
$$
 \eqalign{ {~~}
{\cal Q}({{} U}) ~=~ i y \,(\, 1 - y \,) ~ {\rm Tr} \Big[ ~ {{} 
U}{}^{\dagger}{{} U}{\Hat {\cal G}} \, {\mit\P}_w ~{\cal X} ~ 
\Big] } ~~~,
$$
$$ \eqalign{ {~~~~~}
{\cal X} ~\equiv~ & \, \Big[ ~ {\Hat {\cal G}} \, {\Bar {\mit\P}
}{}^{\,\ad} \, {{} U}^{\dagger} \, {{} U} \,{\Hat {\cal G}}\,
{\mit\P}_{\un  a} \, , \, {{} U}{}^{\dagger} {{} U} {\Hat {\cal 
G}}^2 \,{\mit\P}^{\a}  ~\Big] ~+~ \Big\{ ~ {\Hat {\cal G}} \,
{\Bar {\mit\P}}{}_{\un a} \, {{} U}^{\dagger} \, {{} U} \, {\Hat 
{\cal G}} \, {{\mit\P} }{}^{\,\a} \, , \, {\Hat {\cal H}} ~ \Big\} 
\, {\Hat {\cal  G}} \, {\Bar {\mit\P}}{}^{\ad}  \cr
&{~~~~~}+~ i \, \fracm 12 \, \Big[ \, {\Hat {\cal G}} \,\Big( 
{\Bar D}{}^{\ad} ~ {\Bar {\mit\P}}{}_{\,\ad} \, -\, {\Bar {\mit\P}
}{}^{\ad} {\Bar {\mit\P}}{}_{\, \ad} ~ \Big) \, {{} U}^{\dagger} \, 
{{} U} \, {\Hat {\cal G}}\, {\mit\P}^{\a} \, , \, {{} U}{}^{\dagger
}{{} U}{\Hat {\cal  G}}^2 \,{\mit\P}_{\a} ~ \Big] \cr 
&{~~~~~}+~ i \, \fracm 12 \, \Big\{ \, {\Hat {\cal G}} \,
{\Bar {\mit\P}}{}^{\,\ad} \,  
{{} U}^{\dagger} \, {{} U} \, {\Hat {\cal G}}\, \Big( {D}{}^{\a} ~ {{
\mit\P}}{}_{\,\a}\,-\, {{\mit\P}} {}^{\a}{{\mit\P}}{}_{\,\a} ~ \Big) 
\, , \, {\Hat {\cal H}} \, \Big\} \, {\Hat {\cal G}} \, {\Bar
 {\mit\P}}{}_{\ad} \cr  
&{~~~~~} +~ i \,  (\,1-y\,)\, \Big[  \,{\Hat {\cal G}} \, {\Bar
 {\mit\P}}{}^{\ad} \, {\Hat {\cal G}} \, {\Bar {\mit\P}}{}_{\,\ad} 
\,  {{} U}^{\dagger} \, {{} U} \, {\Hat {\cal G}}\, {\mit\P}^{\a} 
\, , \, {{} U}{}^{\dagger} {{} U} {\Hat {\cal G}}^2 \, {\mit\P
}_{\a} ~ \Big]  \cr
&{~~~~~} +~ i \, (\,1-y\,) \, \Big\{ \, {\Hat {\cal G}} \, {\Bar 
{\mit\P}}{}^{\ad} \,  {{} U}^{\dagger} \,{{} U} \, {\Hat {\cal G
}}\, {{\mit\P}}{}^{\,\a} \,  {\Hat {\cal G}} \,{\mit\P}_{\a} \, , 
\, {\Hat {\cal H}} \, \Big\} \, {\Hat  {\cal G}} \, {\Bar {\mit\P}}
{}_{\ad}  ~~~.}   
\eqno(2.14) $$
Eqs. (2.6, 2.14) represent our explicit superspace form of the
supersymmetric WZNW action.

In order to make a comparison with the action in \cite{NR} we can
introduce the holomorphic ``Lie-group frames'' ${\rm L }_{j}{}^{k
}(\L)$ and ${\rm R}_{j} {}^{k}(\L)$ as well as their anti-holomorphic 
hermitian conjugates via the superfield Maurer-Cartan forms in the
equations
$$ \eqalign{ {~~~~~~~~~~~~}
{{\mit\P}}{}^L &\equiv~ {- i \over f}\,w  (d {\L}^{j} ) \, {\rm 
L}_{j} {}^{ k} \, {t}_{k} ~\,~~,~~~ {\Bar {\mit\P}}{}^L ~\equiv~ 
{i \over f}\,w {{(d {\Bar \L}{}^{j} )}} ~ {\Bar {\rm L}}
{}_{j} {}^{k} \,  {t }_{k} ~~~, \cr
{{\mit\P}}{}^R &\equiv~ {- i \over f}\,w  (d {\L}^{j} ) \, {\rm 
R}_{j} {}^{k} \, {t}_{k} ~~~,~~~ {\Bar {\mit\P}}{}^R ~ \equiv~ 
{i \over f}\, w  {{(d {\Bar \L}{}^{j} )}} ~ {\Bar {\rm R}}{}_{
j} {}^{k} \, {t }_{k} ~~~.}
\eqno(2.15) $$
The symbol ${t}_{k}$ denotes the hermitian group generators.  The 
left-invariant holomorphic frames can be calculated as power series 
from 
$$
\eqalign{ {~~~}
{\rm L}_{j}{}^{\, k}(\L) &=~  (C_2)^{-1}  {\rm Tr}  \Big\{ 
\, {t}^{k} [ g_2(\fracm 12 L_{\L})\,  {t}_{j} \, ] ~ \Big\} 
~~,~~   L_{\L} {t}_{j} ~\equiv~ { i \over f} \, [~ {\L}^{k}
{t}_{k} \, , \,{t}_{j} ~] ~~, ~~ }
\eqno(2.16) $$
where $g_2(x) =  e^{-x} {\rm {sinh}}(x)/ x$ and the constant $C_2$ is
chosen so that ${\rm L}_{j}{}^{\, k}(0) = \d_{j}{}^{k}$.  The
right-invariant  holomorphic frames arise as ${\rm R}_{j}{}^{\, k}(\L) ~=~
{\rm L}_{j }{}^{\, k}(-\L)$.  (If we integrate with respect to a real
variable  the determinant of the Lie-group frames over one copy of the 
group elements, this calculates the volume of the group, for any compact
group.)  By use of (2.15) an alternative form of the 4D, $N$ = 1
supersymmetric ungauged WZNW action is
$$ \eqalign{ 
{\cal S}{}_{WZNW} \,=\,  C_0 \, (\fracm 1{\, 4 \pi^2 \,})\,{\cal R}{\rm
e} \Big\{ \, \int d^8 z& ~ \Big[ ~ {\cal T}_{{i} {j} {\Bar {k}}} 
\, (D^{\a} \L^{i}) \, (\pa_{\un a} \L^{j} ) \, ({\Bar D}{}^{\Dot 
\a}  {\Bar \L}{}^{ ~k}) {~~~~~~~~~~~~~~~~~~~~~~~~~~~} \cr 
&~~+\,{\cal T}_{{i} {\Bar {j}} \, {\Bar {k}} } \, (D^{2} \L^{
i}) \, ({\Bar D}{}^{\Dot \a} \, {\Bar \L}{}^{\, j} ) \, ({\Bar 
D}{}_{ \Dot \a} {\Bar \L}{}^{ ~k})  \cr
&~~+\,{\cal T}_{{i} {j} \, {\Bar {k}} \, {\Bar {h}} } \, (D^{
\a} \L^{i}) \, (D_{\a} \L^{\, j}) \, ({\Bar D}{}^{\Dot \a} 
\, {\Bar \L}{}^{~k} ) \, ({\Bar D}_{\Dot \a}  {\Bar \L}{}^{
~h}) ~\Big] ~ \Big\} ~~~,} \eqno(2.17) $$
where the transcendental coefficients ${\cal T}_{{i} {j} \,
{\Bar {k}} }(\L, \, {\Bar \L})$, ${\cal T}_{{i} {\Bar {j 
}} \, {\Bar{k}} }(\L, \, {\Bar \L})$ and ${\cal T}_{{i} 
{j} \, {\Bar{k}} \, {\Bar{h}}}(\L,\, {\Bar \L})$ can 
be obtained explicitly from (2.14) \cite{GGKP}. \newline

\noindent
{\Large {\bf {(III.) Concluding Remarks
${~~}$}}  }

The next major step to take along the direction of this research is 
to solve the problem of gauging both left and right symmetries in the 
context of the action in (2.14) or (2.17).  This would allow the systematic 
investigation of supersymmetrical electro-weak couplings to the
supersymmetric hadronic model that we have described.  We also note that
the 4D, $N$ = 1 SUSY WZNW model is invariant under the set of variations 
$$
\d \L^{j} ~=~ - \, i \, f \,  [\, 
{\Tilde \a}^{k} ({\rm L}{}^{-1} )_{k} {}^{j} ~-~
{\a}^{k} ({\rm R}{}^{-1} )_{k} {}^{j} ~] ~~~,
\eqno(3.1) $$
with constant real parameters ${\Tilde \a}^{k}$ and ${\a}^{k}$ and 
these transformations correspond to the $SU_L(3) \otimes SU_R(3)$ flavor
symmetries of ordinary QCD.  This has implications for the metric term of
the actual 4D, $N$ = 1 SUSY low-energy QCD model.  Unlike the case of 4D,
$N$ = 2 SUSY YM theory \cite{SW}, there is no {\it {direct}} paradigm for
the non-perturbative structure of the {\it {unbroken}} $N$ = 1 theory.  It
is therefore of some interest to see if the corresponding $N$ = 2 vector 
multiplet or hypermultiplet effective action $\s$-model terms are capable 
of shedding some light on the $N$ = 1 problem.  In particular, it will
be important to see if the $N$ = 1 truncations of the $N$ = 2 models also
possess these symmetries.

A simple examination of the non-supersymmetric WZNW model shows that  
it is on the pure gauge orbit of the appropriate consistent
anomaly\footnote{By way of comparison, the pure gauge orbit of the
covariant anomaly is zero!}.  By its method of construction the action in
(2.6, 2.14) is also on the pure gauge orbit of an anomaly which, as derived
in  the work of \cite{GGP,  GGKPS}, is guaranteed to solve the consistency
condition.   Our result is an exact non-perturbative superfield result for
the  4D, $N$ = 1 Yang-Mills effective action.   Therefore {\it {all}} the 
terms in (2.17) must be included in order to describe the actual 4D, 
$N$ = 1 superspace WZNW action  related to the non-Abelian consistent
anomaly.   Our work also likely implies that there is no choice of the
function $\b_{{i}{j} {\Bar  {k}}}$ in \cite{NR} such that the resulting N-R
WZNW action is on the supersymmetric gauge orbit  of a superspace anomaly
that satisfies the Wess-Zumino consistency condition.   This is because,
although one can  choose $\beta_{{i}{j} \, {\Bar {k}}} = {\cal T}_{{i}{j}
\, {\Bar {k}} }$, there are no {\it {local}} redefinitions that begin
solely with  the N-R WZNW action  and produce the ${\cal T}_{{i} {\Bar {j}}
\, {\Bar {k}}}$ and ${\cal T}_{{i}{\,j} \, {\Bar {k}} \, {\Bar {h}}}$ terms
in (2.17).   

Our work is based on a particular choice ${\cal G}$ of the homotopy 
function used in determining the consistent anomaly.  Although this 
choice is not unique, it has allowed us to obtain an explicit expression 
for the anomaly and the corresponding WZNW action. As shown in ref.\
\cite{GGKPS} any other choice leads to cohomologically equivalent results.
Aside from this, we have introduced two free parameters $f$ and $C_0$.
One free parameter, $f$, appears exactly as in the nonsupersymmetric case
where it corresponds to the pion-decay constant.  The other constant 
$C_0$, which is proportional to the number of QCD colors, is determined by
the normalization in the non-supersymmetric case.  Although our final
expressions are not simple, they can be used for actual superspace
calculations of supersymmetric processes which may ultimately make
experimentally verifiable predictions.

${~~~}$ \newline
${~~~~~~~~~}$``{\it {I resent your insinuendoes.}}'' \newline
${~~~~~~~~~~~}$ -- Richard J. Daley
$${~~~}$$
\noindent



\begin{thebibliography}{66}

\bibitem{HGSX}``Higgs Candidates in e${}^+$e${}^-$ Interactions at 
s = 206.6 GeV,'' L3 Collaboration, M.\ Acciarri, et.\ al., hep-ex/0011043;
``Observation of an Excess in the Search for the Standard Model Higgs
Boson at ALEPH,'' ALEPH Collaboration, R.\ Barate, et.\ al.,
hep-ex/0011045, (http://www.elsevier.nl/NEW/03702693/in${\underline
{~}}$press).

\bibitem{Nonmwznw}S.\ J.\ Gates, Jr., Phys.\ Lett.\ 
{\bf {365B}} (1996) 132; idem.\ Nucl.\ Phys.\ {\bf {B485}} 
(1997) 145; S.\ J.\ Gates, Jr., M.\ T.\ Grisaru, M.\ Ro\v cek, 
O.\ Soloviev and M.\ Knutt-Wehlau, Phys. Lett. {\bf {396B}} 
(1997) 167; S.\ J.\ Gates, Jr.\ and L.\ Rana, Phys.\ Lett.\ 
{\bf {439B}} (1998) 319.

\bibitem{GGP}S.\ J.\ Gates, Jr., M.\ T.\ Grisaru and S.\
Penati, Phys.\ Lett.\ {\bf {481B}} (2000) 397.

\bibitem{GGKPS}S.\ J.\ Gates, Jr., M.\ T.\ Grisaru, M.\ E.\ Knutt, S.\ Penati 
and H.\ Suzuki, ``Supersymmetric Gauge Anomaly with General Homotopic
Paths,'' hep-th/0009192, to appear in Nucl.\ Phys.\ B.

\bibitem{NR}D.\ Nemeschansky and R.\ Rohm, Nucl.\ Phys.\ {\bf {B249}} 
(1985) 157.

\bibitem{Man}A.\ Manohar, Phys.\ Rev.\ Lett.\ {\bf {81}} (1998) 1558.

\bibitem{GGKP} S.\ J.\ Gates, Jr., M.\ T.\ Grisaru, M.\ E.\ Knutt and 
S.\ Penati, in preparation.

\bibitem{SW}N.\ Seiberg and E.\ Witten, Nucl.\ Phys.\ {\bf B426} (1994) 19.


\end{thebibliography}
\end{document}